\documentstyle[preprint,aps,graphicx]{revtex}

\tightenlines
\newlength{\Mlength}\Mlength 12cm

\begin{document}
\title{Ab intio calculation of the Transmission Coefficients from a Superlattice
electronic structure}
\author{Ingmar Riedel, Peter Zahn, and Ingrid Mertig}
\address{Institut f\"{u}r Theoretische Physik\\
Technische Universit\"{a}t Dresden, D-01062 Dresden, Germany}
\date{\today}
\maketitle

\begin{abstract}
A new formalism to calculate the transmission coefficient $t$ of electrons
from a material ${\cal L}$ into the same material ${\cal L}$ through a
barrier ${\cal B}$ is presented. The barrier ${\cal B}$ is arbitrary and can
be metallic, semiconducting or insulating. The important feature of this
formalism is that it starts from the electronic structure of a periodic $%
{\cal L}$/${\cal B}$ superlattice. The electronic stucture of such a
superlattice is calculated selfconsistently within a Screened
Korringa-Kohn-Rostoker method. The capability of the new method is
demonstrated by means of a free electron model. First results for the
transmission of Cu electrons through a Co layer are presented.
\end{abstract}

\section{Introduction}

If Bloch electrons encounter planar interfaces they may be reflected or
transmitted. In this paper we develop a technique for the calculation of
these reflection and transmission amplitudes within the Screened
Korringa-Kohn-Rostoker (Screened KKR) approach, a technique for ab initio
electronic structure calculations.

There are many uses for Bloch wave transmission and reflection amplitudes
and probabilities. First of all they can be applied to describe transport 
\cite{c:Landauer70,c:Fischer} in films and multilayers including the
electronic structure of the real material. Secondly, they can be used do
develop a first principle based theory of electron tunneling. Finally,
transmission and reflection coefficients are the basic ingredients for the
theoretical understanding of interlayer exchange coupling \cite{c:Bruno95}.

Consider an arbitrary barrier ${\cal B}$ which is sandwiched by two leads of
the material ${\cal L}$ (see Fig. 1). For the propagating bulk-states in $%
{\cal L}$ the transmission coefficient $t$ through the barrier is defined as 
\begin{equation}
t=\frac{A_{1}}{A_{0}}\text{\ \ for }B_{1}=0  \label{eg:def_t}
\end{equation}
where the $A_{0,1}$ and $B_{1}$ are the complex amplitudes of the right- and
left-moving Bloch waves. During the scattering process the energy $E$ and
the in-plane component ${\bf k}_{\parallel }$ of the wave vector $(k,{\bf k}%
_{\parallel })$ are conserved.

In the past different methods were established for the calculation of $t$,
e.g. \cite{c:Stiles96,c:MacLaren99,c:Wildberger98}. These
methods usually model the structure of the two semi-infinite leads ${\cal L}$
separated by a barrier ${\cal B}$. Via the Green's function or the wave
function of the system $t$ can be computed. In this work we present a new
method which approaches the problem differently: We use the band structure
and wave function of a periodic superlattice ${\cal B}$/${\cal L}$ to
calculate $t$. This is strongly motivated by the development of highly
efficient N-scaling superlattice KKR codes, e.g. 
\cite{c:Zeller95,c:Zahn97,c:Anderson84,c:Szunyogh94}.

A superlattice is a structure ${\cal B}$/${\cal L}$ which is periodically
repeated in all three dimensions (see Fig. 2). Therefore the case of the
single barrier ${\cal B}$ with leads ${\cal L}$ is fully contained in this
structure and the similarity with the Kronig-Penney model is obvious. Kronig
and Penney showed in their simple 1-dimensional free electron model how the
bandstructure of a periodic system can be derived from the scattering
properties of a single barrier. (For more details on the Kronig-Penney model
see for instance \cite{c:Merzbacher70}.)

The keynote of this paper is to derive the scattering properties of a single
barrier from the electronic structure of a superlattice. We discuss the
theory and numericalimplementation of the new method and present its first
successful applications to free electrons and a Cu/Co (001) system.

\section{Theory}

First we state our basic assumptions. Thereafter we discuss the transmission
coefficient $t$ of the single barrier and finally its connection to the
superlattice wave function.

\subsection{Basic assumptions}

In Fig. 1 and Fig. 2 schematic pictures of a single barrier and a
superlattice are presented. Note that $x$ is always perpendicular to the
barrier ${\cal B}$, the other two components are denoted with ${\bf r}%
_{\parallel }$. Along ${\bf r}_{\parallel }$ the leads ${\cal L}$ and the
barrier ${\cal B}$ show the same 2-dimensional periodicity. The
superlattice, however, is also periodic in $x$-direction.

The single barrier case (see Fig. 1) consists of the barrier and two
semi-infinite leads. In difference to the common understanding, we will
include the perturbed interface regions of the leads into the barrier ${\cal %
B}$. The leads ${\cal L}$, however, consist of bulk-like potentials only. A
superlattice (see Fig. 2) consists of an infinite number of barriers
separated by interlayers. The barriers ${\cal B}$ contain again the
perturbed interface regions. The interlayers consist of bulk-like
potentials, the same as in the leads ${\cal L}$ of the single barrier case.

Furthermore, we assume energy- and ${\bf k}_{\parallel }$-conservation due
to the elastic scattering and the in-plane symmetry of the system. Since $%
{\bf k}_{\parallel }$ is conserved the problem can be treated as quasi
1-dimensional, but the reader should be cautious while keeping track of the
components of the other two dimensions.

We make two additional simplifications: First, the barrier is symmetric and,
secondly, we restrict our considerations to one propagating bulk state $\Phi
^{(k,{\bf k}_{\parallel })}(x,{\bf r}_{\parallel })$ at a given energy and $%
{\bf k}_{\parallel }$. That means, degeneration and band-crossing of bulk
states in the leads are excluded.

\subsection{The single barrier}

The wavefunction in the leads ${\cal L}$ is a superposition of the right-
and left-moving bulk states, since the solution of the Schr\"{o}dinger
equation depends only on the local potential (see Fig. 1): 
\begin{equation}
\Psi (x,{\bf r}_{\parallel })=A_{n}e^{ikx}e^{i{\bf k}_{\parallel }{\bf r}%
_{\parallel }}u_{k,{\bf k}_{\parallel }}(x,{\bf r}_{\parallel
})+B_{n}e^{-ikx}e^{i{\bf k}_{\parallel }{\bf r}_{\parallel }}u_{-k,{\bf k}%
_{\parallel }}(x,{\bf r}_{\parallel })\text{ .}  \label{eq:WaveSingle}
\end{equation}
$n=\left\{ 0,1\right\} $ denotes the two sides of the barrier. $\Phi ^{(k,%
{\bf k}_{\parallel })}(x,{\bf r}_{\parallel })=e^{ikx}e^{i{\bf k}_{\parallel
}{\bf r}_{\parallel }}u_{k,{\bf k}_{\parallel }}(x,{\bf r}_{\parallel })$ is
the Bloch function of the bulk system. The decomposition of $\Phi ^{(k,{\bf k%
}_{\parallel })}(x,{\bf r}_{\parallel })$ into a phase factor and a periodic
function is due to Bloch's theorem with $u_{k,{\bf k}_{\parallel }}(x,{\bf r}%
_{\parallel })=u_{k,{\bf k}_{\parallel }}(x-md,{\bf r}_{\parallel })$, where 
$d$ is the bulk lattice parameter and $m$ an integer.

The amplitudes $A_{n}$ and $B_{n}$ on both sides of the barrier are related
to each other by a transfer matrix $M$ which describes all the scattering
processes inside the barrier: 
\begin{equation}
\left( 
\begin{array}{c}
A_{0} \\ 
B_{0}
\end{array}
\right) =M\left( 
\begin{array}{c}
A_{1} \\ 
B_{1}
\end{array}
\right) \text{.}
\end{equation}

The four complex variables of this $M$-matrix are not independent from each
other. Due to time-reversal invariance, current conservation and the
symmetry of the barrier, $M$ can be parametrized with two independent real
parameters $\alpha _{1}$ and $\beta _{1}$ in the following form \cite
{c:Merzbacher70}: 
\begin{eqnarray}
M &=&\left( 
\begin{array}{cc}
\alpha _{1}+i\beta _{1} & i\beta _{2} \\ 
-i\beta _{2} & \alpha _{1}-i\beta _{1}
\end{array}
\right)  \\
\text{with \ \ \ \ \ \ }\beta _{2}^{2} &=&\alpha _{1}^{2}+\beta _{1}^{2}-1%
\text{ .}  \nonumber
\end{eqnarray}
Therefore the relation $M^{-1}=M^{\ast }$ is valid.

Using the components of the transfer matrix the transmission coefficient $t$
is by definition Eq. (\ref{eg:def_t}): 
\begin{equation}
t=\frac{1}{\alpha _{1}+i\beta _{1}}\text{ .}
\end{equation}

Obviously, the problem is solved if the eigenstates $\Psi (x,{\bf r}%
_{\parallel })$ of such a single barrier are known.

\subsection{The superlattice}

In contrast to the eigenfunctions of the single barrier (see Eq. (\ref
{eq:WaveSingle})) the eigenfunctions of a superlattice $\Psi (x,{\bf r}%
_{\parallel })$ are described by quantum numbers $(\kappa ,{\bf k}%
_{\parallel })$. Here, $\kappa $ is the superlattice wave vector in $x$%
-direction reflecting the prolonged lattice structure with the lattice
constant $l$, that is the length of the unit cell in $x$-direction.
Furthermore, the in-plane component ${\bf k}_{\parallel }$ is the same as in
the bulk system since in-plane symmetry is conserved. The translational
symmetry of $\Psi ^{(\kappa ,{\bf k}_{\parallel })}$ in the superlattice
direction is given by Bloch's theorem: 
\begin{equation}
\Psi ^{(\kappa ,{\bf k}_{\parallel })}(x,{\bf r}_{\parallel })=\left(
e^{-i\kappa l}\right) ^{n}\Psi ^{(\kappa ,{\bf k}_{\parallel })}(x+nl,{\bf r}%
_{\parallel })\text{ ,}  \label{eg:Bloch}
\end{equation}
where the integer $n$ ranges from minus to plus infinity.

Let us now consider the superlattice as a periodic repetition of single
barriers ${\cal B}$ with the interlayer ${\cal L}$ in between. In the
interlayer the potential is bulk-like and therefore we can expand the
superlattice wave function (see Eq. (\ref{eg:Bloch})) as a superposition of
right- and left-moving Bloch functions of the bulk system in the cell $n=0$ $%
(a\leq x\leq a+s)$: 
\begin{eqnarray}
&&A_{0}e^{ikx}e^{i{\bf k}_{\parallel }{\bf r}_{\parallel }}u_{k,{\bf k}%
_{\parallel }}(x,{\bf r}_{\parallel })+B_{0}e^{-ikx}e^{i{\bf k}_{\parallel }%
{\bf r}_{\parallel }}u_{-k,{\bf k}_{\parallel }}(x,{\bf r}_{\parallel })
\label{eq:decomp} \\
&=&\left( e^{-i\kappa l}\right) ^{n}\left[ A_{n}e^{ik(x+nl)}e^{i{\bf k}%
_{\parallel }{\bf r}_{\parallel }}u_{k,{\bf k}_{\parallel }}(x+nl,{\bf r}%
_{\parallel })+B_{n}e^{-ik(x+nl)}e^{i{\bf k}_{\parallel }{\bf r}_{\parallel
}}u_{-k,{\bf k}_{\parallel }}(x+nl,{\bf r}_{\parallel })\right] \text{ .} 
\nonumber
\end{eqnarray}
By using the relation $u_{k,{\bf k}_{\parallel }}(x+nl,{\bf r}_{\parallel })=
$ $u_{k,{\bf k}_{\parallel }}(x,{\bf r}_{\parallel })$ we can rewrite this
into 
\begin{eqnarray}
&&A_{0}e^{ikx}e^{i{\bf k}_{\parallel }{\bf r}_{\parallel }}u_{k,{\bf k}%
_{\parallel }}(x,{\bf r}_{\parallel })+B_{0}e^{-ikx}e^{i{\bf k}_{\parallel }%
{\bf r}_{\parallel }}u_{-k,{\bf k}_{\parallel }}(x,{\bf r}_{\parallel }) \\
&=&\left( e^{-i\kappa l}\right) ^{n}\left[ A_{n}e^{iknl}e^{ikx}e^{i{\bf k}%
_{\parallel }{\bf r}_{\parallel }}u_{k,{\bf k}_{\parallel }}(x,{\bf r}%
_{\parallel })+B_{n}e^{-iknl}e^{-ikx}e^{i{\bf k}_{\parallel }{\bf r}%
_{\parallel }}u_{-k,{\bf k}_{\parallel }}(x,{\bf r}_{\parallel })\right] 
\text{ .}
\end{eqnarray}
Since the right- and left-moving bulk states are orthogonal we find 
\begin{equation}
\left( 
\begin{array}{c}
A_{0} \\ 
B_{0}
\end{array}
\right) =\left( e^{-i\kappa l}\right) ^{n}\left( 
\begin{array}{c}
A_{n}e^{iknl} \\ 
B_{n}e^{-iknl}
\end{array}
\right) \text{ .}  \label{eg:com1}
\end{equation}
On the other hand we can connect the amplitudes in different cells with the
transfer matrix $M$: 
\begin{equation}
\left( 
\begin{array}{c}
A_{0} \\ 
B_{0}
\end{array}
\right) =M^{n}\left( 
\begin{array}{c}
A_{n} \\ 
B_{n}
\end{array}
\right) \text{ .}  \label{eq:com2}
\end{equation}

We can combine Eqs. (\ref{eg:com1}) and (\ref{eq:com2}) to the eigenvalue
equation 
\begin{equation}
e^{i\kappa l}\left( 
\begin{array}{c}
A_{0} \\ 
B_{0}
\end{array}
\right) =P\left( 
\begin{array}{c}
A_{0} \\ 
B_{0}
\end{array}
\right) 
\end{equation}
with 
\begin{eqnarray}
P &=&\left( 
\begin{array}{cc}
e^{ikl} & 0 \\ 
0 & e^{-ikl}
\end{array}
\right)M^{-1} \\
&=&\left( 
\begin{array}{cc}
(\alpha _{1}-i\beta _{1})e^{ikl} & -i\beta _{2}e^{ikl} \\ 
i\beta _{2}e^{-ikl} & (\alpha _{1}+i\beta _{1})e^{-ikl}
\end{array}
\right) \text{ .}
\end{eqnarray}

Since $P$ is fully determined by the single eigenvalue $e^{i\kappa l}$ and
the corresponding eigenvector $\left( 
\begin{array}{c}
A_{0} \\ 
B_{0}
\end{array}
\right) $, we obtain the following algebraic set for $\alpha _{1}$ and $%
\beta _{1}$: 
\begin{eqnarray}
\beta _{2}^{2} &=&\alpha _{1}^{2}+\beta _{1}^{2}-1  \nonumber \\
\frac{A_{0}}{B_{0}} &=&\frac{\beta _{2}e^{ikl}}{\alpha _{1}\sin kl-\beta
_{1}\cos kl-\sin \kappa l}  \label{eg:2} \\
\cos \kappa l &=&\alpha _{1}\cos kl+\beta _{1}\sin kl\text{ .}  \nonumber
\end{eqnarray}
Thus, we are able to calculate $t$ for a given energy and ${\bf k}%
_{\parallel }$ knowing $k$, $\kappa $ and $\frac{A_{0}}{B_{0}}$ of the
superlattice wave function $\Psi ^{(\kappa ,{\bf k}_{\parallel })}$ and
corresponding bulk state $\Phi ^{(k,{\bf k}_{\parallel })}$.

\section{Numerical aspects}

In this section we explain how to extract $k$ and $\frac{A_{0}}{B_{0}}$ from 
$\Psi ^{(\kappa ,{\bf k}_{\parallel })}(x,{\bf r}_{\parallel })$ in the
interlayer, which is very similar to a Fourier analysis of the $\Psi
^{(\kappa ,{\bf k}_{\parallel })}(x,{\bf r}_{\parallel })$. For the
electronic band structure calculations we used a Screened
Korringa-Kohn-Rostoker (Screened KKR) method.

\subsection{Brief introduction to the Screened KKR method}

The Screened KKR method (see 
\cite{c:Zeller95,c:Zahn97,c:Anderson84,c:Szunyogh94}) was developed for the selfconsistent
calculation of the electronic structure. It is based on Green's functions
and the computation time scales linearly with the number of atoms in the
unit cell. This allows the computation of systems with a relatively large
number of atoms. For the atomic potentials the atomic sphere approximation
(ASA) is used. The superlattice wave function is expanded in terms of
spherical solutions: 
\begin{equation}
\Psi ^{{\bf \kappa }}({\bf R}^{n}+{\bf r}_{\mu }+{\bf r})=%
\mathrel{\mathop{\sum }\limits_{L}}%
e^{i{\bf \kappa R}^{n}}C_{L}^{\mu }({\bf \kappa )}R_{l}^{\mu }(r,E)Y_{L}(%
\widehat{r})\text{ ,}  \label{eg:defofcompcoeff}
\end{equation}
the wave function in the leads as well: 
\begin{equation}
\Phi ^{{\bf k}}({\bf R}^{n}+{\bf r}_{\mu }+{\bf r})=%
\mathrel{\mathop{\sum }\limits_{L}}%
e^{i{\bf kR}^{n}}\widehat{C}_{L}^{\mu }({\bf k)}R_{l}^{\mu }(r,E)Y_{L}(%
\widehat{r})\text{ .}  \label{eq:def2}
\end{equation}

${\bf R}^{n}$ is the lattice vector pointing to the $n^{th}$ unit cell, $%
{\bf r}_{\mu }$ refers to the $\mu ^{th}$ atom in the unit cell and ${\bf r}$
is the space coordinate inside the $\mu ^{th}$ atomic sphere. The $Y_{L}(%
\widehat{r})$ are the spherical harmonics with the short hand notation $%
L=(l,m)$, $l$ is the angular momentum and $m$ the magnetic quantum number. $%
R_{l}^{\mu }(r,E)$ is the radial solution of the Schr\"{o}dinger equation in
the $\mu ^{th}$ atomic sphere. The $C_{L}^{\mu }({\bf \kappa )}$ and $%
\widehat{C}_{L}^{\mu }({\bf k)}$ are the expansion coefficients, ${\bf %
\kappa =(}\kappa {\bf ,k_{\parallel })}$ and ${\bf k=}(k{\bf ,k_{\parallel })%
}$ are the Bloch vectors in the superlattice and in the bulk system,
respectively.

Angular momenta up to $l_{\max }=3$ were taken into account.

\subsection{Implementation}

As already explained, the superlattice wave function $\Psi ^{{\bf (}\kappa 
{\bf ,k_{\parallel })}}$ in the interlayer ${\cal L}$ can be decomposed in
right- and left-moving Bloch waves $\Phi ^{(k{\bf ,k_{\parallel })}}$ and $%
\Phi ^{(-k{\bf ,k_{\parallel })}}$: 
\begin{equation}
\Psi ^{{\bf (}\kappa {\bf ,k_{\parallel })}}=A_{0}\Phi ^{(k{\bf %
,k_{\parallel })}}+B_{0}\Phi ^{(-k{\bf ,k_{\parallel })}}\text{ .}
\label{eq:decompos}
\end{equation}
Using the angular momentum expansions of Eqs. (\ref{eg:defofcompcoeff}) and (%
\ref{eq:def2}) we can expand Eq. (\ref{eq:decompos}) in the $0^{th}$ unit
cell, which corresponds to ${\bf R}^{0}=(-l,0,0)$: 
\begin{eqnarray}
&&%
\mathrel{\mathop{\sum }\limits_{L}}%
e^{-i\kappa l}C_{L}^{\mu }(\kappa {\bf ,k_{\parallel })}R_{l}^{\mu
}(r,E)Y_{L}(\widehat{r})  \nonumber \\
&=&%
\mathrel{\mathop{\sum }\limits_{L}}%
\left[ 
\begin{array}{c}
A_{0}e^{-ikl}\widehat{C}_{L}^{\mu }(k{\bf ,k_{\parallel })}R_{l}^{\mu
}(r,E)Y_{L}(\widehat{r}) \\ 
+B_{0}e^{ikl}\widehat{C}_{L}^{\mu }(-k,{\bf k}_{\parallel }{\bf )}R_{l}^{\mu
}(r,E)Y_{L}(\widehat{r})
\end{array}
\right] \text{ .}
\end{eqnarray}
Now we make use of the orthogonality relation of the $Y_{L}(\widehat{r})$
and the following relations between the $\widehat{C}_{L}^{\mu }(\pm k{\bf %
,k_{\parallel })}$ due to the real potential of the leads and Bloch's
theorem: 
\begin{eqnarray}
\widehat{C}_{L}^{\mu }(\pm k{\bf ,k_{\parallel })} &=&e^{i{\bf (\pm }k{\bf %
,k_{\parallel }})(r_{{ \mu }}{\bf -r}_{0}{\bf )}}\widehat{C}_{L}^{0}(\pm
k,{\bf k}_{\parallel }{\bf )} \\
\widehat{C}_{L}^{0}(-k,{\bf k}_{\parallel }{\bf )} &=&{\bf (-}1)^{l+m}%
\widehat{C}_{L}^{0}(k,{\bf k}_{\parallel }{\bf )}\text{ .}
\end{eqnarray}
So finally we conclude: 
\begin{eqnarray}
0 &=&e^{-i\kappa l}C_{L}^{\mu }(\kappa {\bf ,k_{\parallel })}
\label{eq:algset} \\
&&-A_{0}e^{-ikl}e^{i{\bf (}k{\bf ,k_{\parallel }})(r_{{ \mu }}{\bf -r}_{0}%
{\bf )}}\widehat{C}_{L}^{0}(k{\bf ,k_{\parallel })}  \nonumber \\
&&-B_{0}e^{ikl}e^{i{\bf (-}k{\bf ,k_{\parallel }})(r_{{ \mu }}{\bf -r}_{0}%
{\bf )}}\widehat{C}_{L}^{0}(k,{\bf k}_{\parallel }{\bf )(-}1)^{l+m}\text{ .}
\nonumber
\end{eqnarray}

We can exploit this equation at three different atomic positions $\mu $ in
the interlayer and receive an algebraic set for $k$ and the ratio $\frac{%
A_{0}}{B_{0}}$. It turns out that the $\widehat{C}_{L}^{0}(k{\bf %
,k_{\parallel })}$ cancel, so no additional computation of the bulk
eigenstates of the leads is necessary. All we have to compute are the
eigenstates of the superlattice $\Psi ^{(\kappa ,{\bf k}_{\parallel })}(x,%
{\bf r}_{\parallel })$ for a given energy $E$ and in-plane wave vector ${\bf %
k}_{\parallel }$ which gives us the $C_{L}^{\mu }(\kappa {\bf ,k_{\parallel
})}$ and $\kappa $. From here on we can calculate $t$ via Eqs. (\ref{eg:2})
and (\ref{eq:algset}). As a side effect we obtain the bandstructure $k(E)$
of the lead material. This gives us the nice opportunity for an intermediate
test of our method as we will see in the next section.

Still there is another problem we are faced with: If $\kappa $ is complex
the corresponding $\Psi ^{(\kappa ,{\bf k}_{\parallel })}(x,{\bf r}%
_{\parallel })$ is evanescent which means it does not exist in the
superlattice. In other words we are in a superlattice bandgap. But also in
this energy range we want to calculate the corresponding $k$ and $t$. One
opportunity is the computation of the complex bandstructure and developing a
similar formalism as the one described above. But usually the complex
bandstructure is not available. Another nice idea is to vary the interlayer
thickness $s=(l-2a)$ (see Fig. 2): The superlattice bandstructure gets
stretched or squeezed on the energy axis comparable to the eigenstates of a
square well. In this way we can scan the whole energy range, albeit it makes
more than one superlattice calculation necessary. We followed this second
idea.

\section{Results}

The results we discuss in this section demonstrate that our new method is
working successfully. We considered the scattering of free electrons through
a rectangular barrier, which is analytically solvable. Furthermore we
investigated the transmission of Cu electrons through 5 monolayers (ML) Co
(001), which we compared with the results of Wildberger \cite{c:Wildberger97}%
, who developed a different method for the computation of $t$. In addition
to the $\left| t\right| (E)$-dependence we checked the dispersion relation $%
k(E)$ of the lead material.

\subsection{Free electrons through a rectangular barrier}

Outside the barrier we modelled an exact free electron gas with constant
potential $V=0$. The barrier itself consists of five layers with muffin tin
(MT) spheres of a constant potential inside the sphere. The potential height
is $V=0.125Ryd$ and they are arranged in an fcc lattice with a lattice
constant of $d=6.76a.u.$. This is only an approximate rectangular barrier,
but the best one possible with ASA potentials.

In Fig. 3 the dependence of $\left| t\right| $ on $(E-{\bf k}_{\parallel
}^{2})$ for ${\bf k}_{\parallel }=\left( \frac{\pi }{2d},\frac{\pi }{2d}%
\right) $ is plotted. The analytical curve was fitted to the numerical data
with a potential height of the rectangular barrier $V=0.76\cdot 0.125Ryd$.
This potential height is given by the dashed line. The factor $0.76$ is very
close to $0.74$ which is the effective space filling of MT spheres on an fcc
lattice. The remaining small difference of $0.02$ and also the subtle
deviations in the $\left| t\right| (E)$-dependence can be understood very
well in terms of the different barrier shapes. The two numerical curves with
the different interlayer thicknesses (circle and plus) coincide, which is a
brilliant test for the validity of the method itself. The discontinuity of
the curves is due to the superlattice bandgaps. Obviously these gaps can be
filled with different interlayer thicknesses $s$. To fill all gaps more
calculations with different interlayer thicknesses $s$ would be necessary.
This is not done here since we just want to demonstrate the capability of
our new method. For a short interpretation we notice that the electrons have
a low transmission coefficient below the barrier since they have to tunnel
through. Above the barrier height they approach $t=1$ very rapidly, the
oscillations are due to interference effects on the barrier interfaces.

In Fig. 4 we show the correspondence between the calculated dependence of $k$
on $(E-{\bf k}_{\parallel }^{2})$ for different ${\bf k}_{\parallel }$
(circle, cross and triangle) and the expected free electron dispersion
(solid line), and obtain a very good agreement.

\subsection{Cu electrons through 5 ML of Co (001)}

We arranged the Co and Cu on a fcc lattice with a lattice constant of $%
6.76a.u.$. The value is larger than that of magnetic fcc Co and smaller than
that of fcc Cu. The Co layer consists of 5 ML. The potentials of the Co
layer and three Cu layers at both sides of the barrier are taken from a
self-consistent Co$_{5}$Cu$_{7}$ supercell calculation and form the barrier $%
{\cal B}$. The potential in the middle of the Cu layer is a bulk-like Cu ASA
potential. This approximation has no influence on the results and has the
advantage that we can construct easily any Cu interlayer thickness $s$
without doing an extra selfconsistent calculation.

Figure 5b and 5d show the $\left| t\right| (E)$-dependence for ${\bf k}%
_{\parallel }=0$ in the majority and minority channels. Since Co is a
ferromagnetic material the superlattice bandstructure of Cu/Co has no more
spin-degeneracy and we obtain spin-splited majority and minority bands. As
for the free electron results the curves for different interlayer
thicknesses $s$ match, which is not shown explicitly here. Large parts of
the $\left| t\right| (E)$-curve can be justified qualitatively by the band
structure (BS)\ of Co and Cu, which are plotted in Fig. 5a (Co majority BS),
Fig. 5c (Cu BS) and Fig. 5e (Co minority BS). First we discuss the $\left|
t\right| (E)$-dependence in the majority channel (see Fig. 5b):

Since the Co and Cu states have the same symmetry character in the energy
region {\bf I} the Cu electrons can propagate through the barrier very well,
below the bottom of the Co band they have to tunnel. Therefore the behavior
in section {\bf I} is very similar to the free electron result as discussed
in the previous subsection. In region {\bf II} there are no propagating Cu
bulk states and consequently there is no $t$. In region {\bf III} we see
that $t$ is almost free electron like again, the rapid breakdown at about
0.45 Ryd corresponds the change of the symmetry character in the Co bands.
Region {\bf IV} has to be excluded from our discussion since the Cu band
structure is degenerated there and hence our method is not applicable. In
region {\bf V}, including the Fermi level at $0.68 Ryd$, we obtain a $t$ of
almost $1$ which is due to the corresponding band symmetries in Co and Cu.

In the minority channel (see Fig. 5d) we see two major differences to the
majority channel in region {\bf III} and {\bf V}: In region {\bf III} $t$ is
almost zero since the Co minority band has a gap there and so no band
matching is possible between the Cu and Co states. In region {\bf V} we see
a low $t$ below 0.64 Ryd. This is due to the upward shifted Co bands in
comparison to the majority channel and therefore no band matching between
states with the same symmetry\ character occurs.

Furthermore, we compare our results in the energy region {\bf V} (see Fig.
6a) with the results obtained with a different method by Wildberger \cite
{c:Wildberger97} (see Fig. 6b), who used a semi-infinite structure with two
Co barriers at a certain distance (which is comparable to our interlayer
thickness $s$) for the calculation of the reflection coefficient $\left|
r\right| $. The reflection coefficient $\left| r\right| $ is related to $%
\left| t\right| $ by $\left| r\right| =\sqrt{1-\left| t\right| ^{2}}$. In
Fig. 6b we see their $\left| r\right| (E)$-dependence for the majority- and
minority-channel (dashed and solid line) for ${\bf k}_{\parallel }=0$ around
the Fermi energy. The agreement with our own results shown in Fig. 6a is
very good. Slight differences are related to the different lattice
parameters which were used. In addition, Wildberger evaluates the Green's
function of the system for energies with a small imaginary part, which
causes a broadening corresponding to a finite temperature. The agreement
between our calculation and the calculation of Wildberger is a confirmation
of the capability and the validity of our new method.

In Fig. 7 we see the band structure of fcc Cu along the line $\Gamma -X$.
The solid line is taken from a pure Cu band structure calculation, the
crosses denote the band structure $k(E)$ extracted from the superlattice
calculation. The region between the dashed lines corresponds to region {\bf %
IV} in Fig. 5 and has to be excluded for the reasons mentioned above. Again,
the correspondence between both curves is perfect.

\section{Summary}

We presented a new method to calculate the transmission coefficient $t$ of
bulk states through a single barrier from a superlattice band structure. Its
validity is fully confirmed by the successful test results. At this state
the method is restricted to symmetric barriers and to only one propagating
bulk state at a given energy and in plane wave vector. The generalization to
the case of mor than one propagating bulk state and unsymmetric barriers is
under development.

\section{Acknowledgment}

We like to thank J. Opitz, J. Binder, M. Turek, W. John (TU Dresden), the
group of G.E.W. Bauer (TU Delft) and the group of P.H. Dederichs
(Forschungszentrum J\"{u}lich) for helpful discussions.

\newpage

\begin{figure}
\begin{center}
  \includegraphics[width=\Mlength,clip=true]{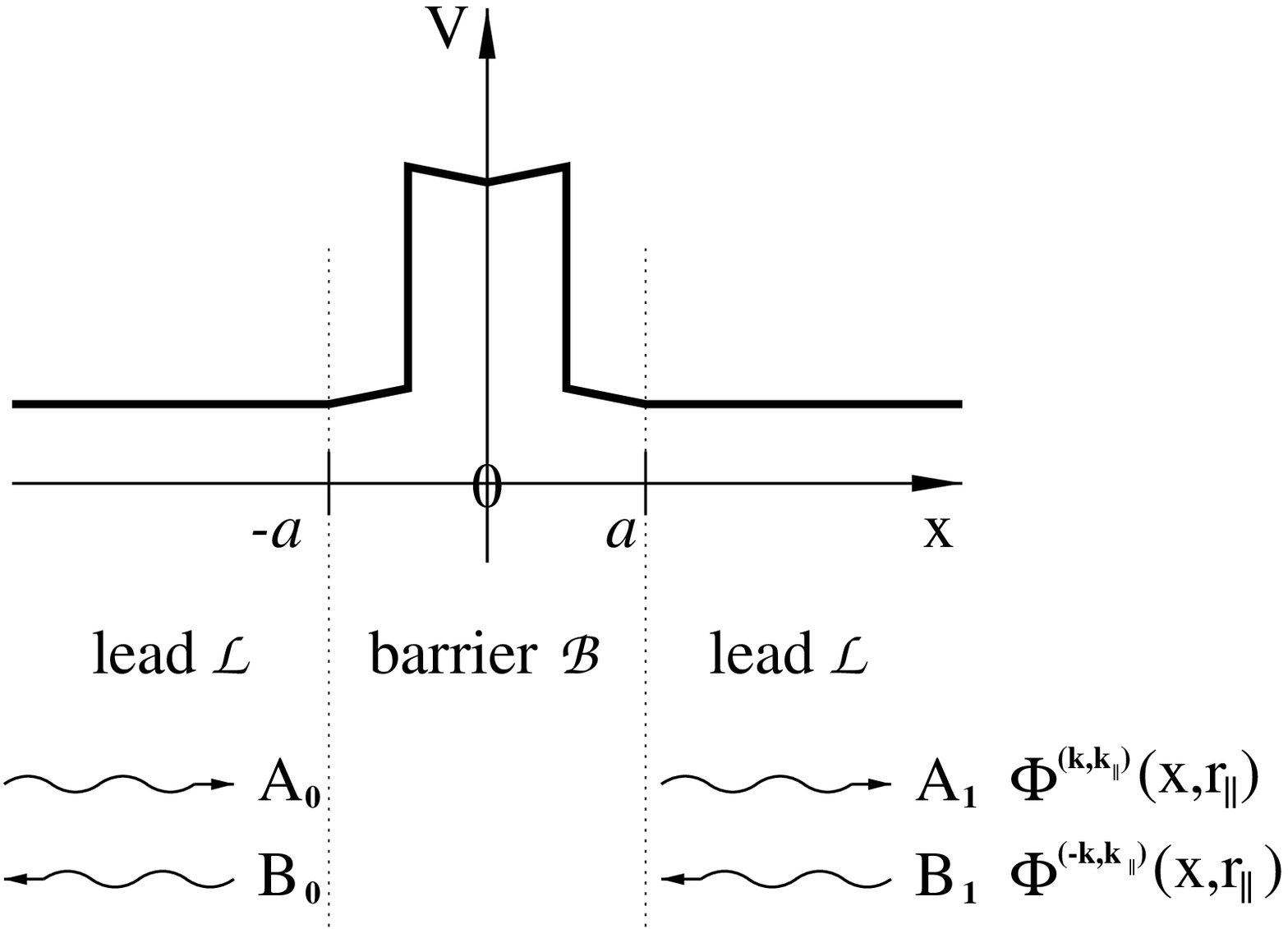}%
\end{center}
\vspace*{1cm}
  \caption{
Schematic drawing of the single barrier case with a barrier ${\cal B}
$ (including the perturbed interface regions) and leads ${\cal L}$.
Furthermore, right- and left-moving eigenstates $\Phi $ are shown.
  }
\end{figure}
\newpage
%
\begin{figure}
\begin{center}
  \includegraphics[width=\Mlength,clip=true]{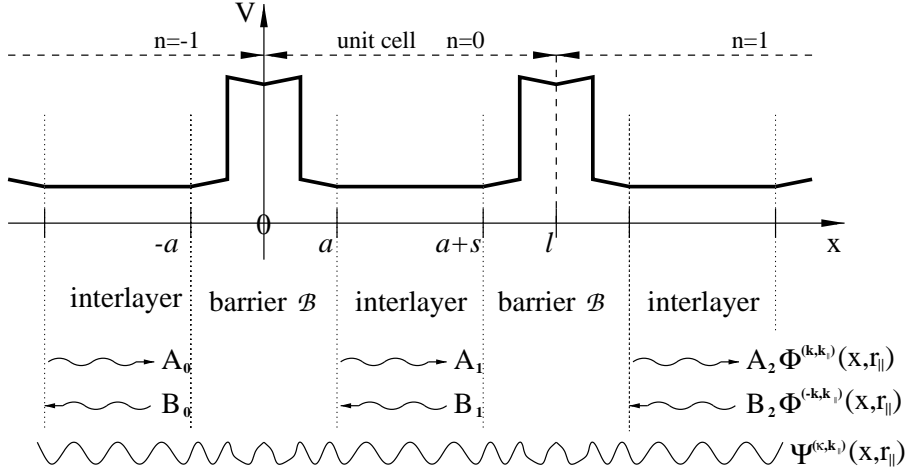}%
\end{center}
\vspace*{1cm}
  \caption{
Schematic drawing of the superlattice with barriers ${\cal B}$
(including the perturbed interface regions) and interlayer ${\cal L}$. The
superlattice wave function $\Psi $ in the interlayers is a superposition of
right- and left-moving bulk eigenstates $\Phi $.
  }
\end{figure}
\newpage
%
\begin{figure}
\begin{center}
  \includegraphics[width=\Mlength,clip=true]{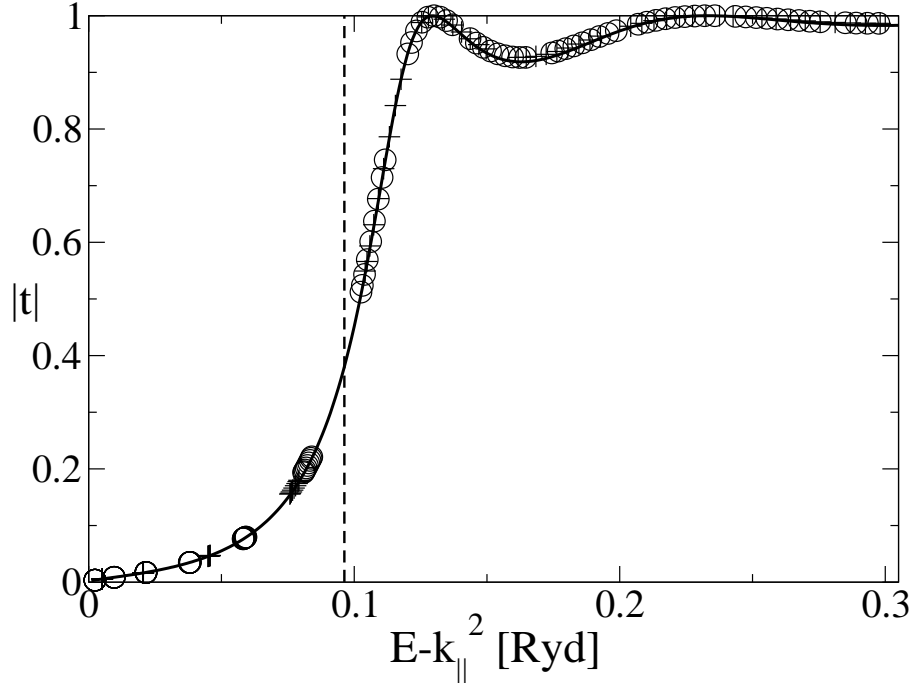}%
\end{center}
\vspace*{1cm}
  \caption{
The dependence of $\left| t\right| $ on $(E-{\bf k}_{\parallel
}^{2}) $ for ${\bf k}_{\parallel }=\left( \frac{\pi }{2d},\frac{\pi }{2d}%
\right) $ of free electrons through a rectangular barrier with a thickness of 5
ML and an averaged height of 0.092 Ryd denoted by the dashed line: Analytical
results (solid line) in comparison with the results obtained with our new
method for two different interlayer thicknesses (cross: 11\ ML, circle: 17
ML).
  }
\end{figure}
\newpage
%
\begin{figure}
\begin{center}
  \includegraphics[width=\Mlength,clip=true]{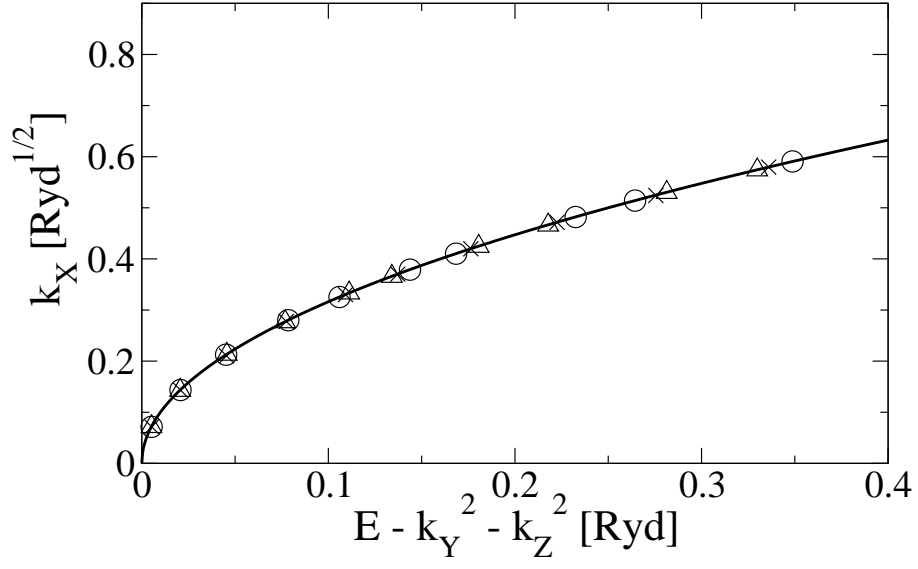}%
\end{center}
\vspace*{1cm}
  \caption{
The dispersion relation of free electrons for different ${\bf k}%
_{\parallel }$ (circle: $k_{x}=k_{y}=0$, cross: $k_{x}=0.5$, $k_{y}=0$;
triangle: $k_{x}=k_{y}=0.5$, all in relative units of the Brillouin zone) in
comparison with the analytical result (solid line).
  }
\end{figure}
\newpage
%
\begin{figure}
\begin{center}
  \includegraphics[width=\Mlength,clip=true]{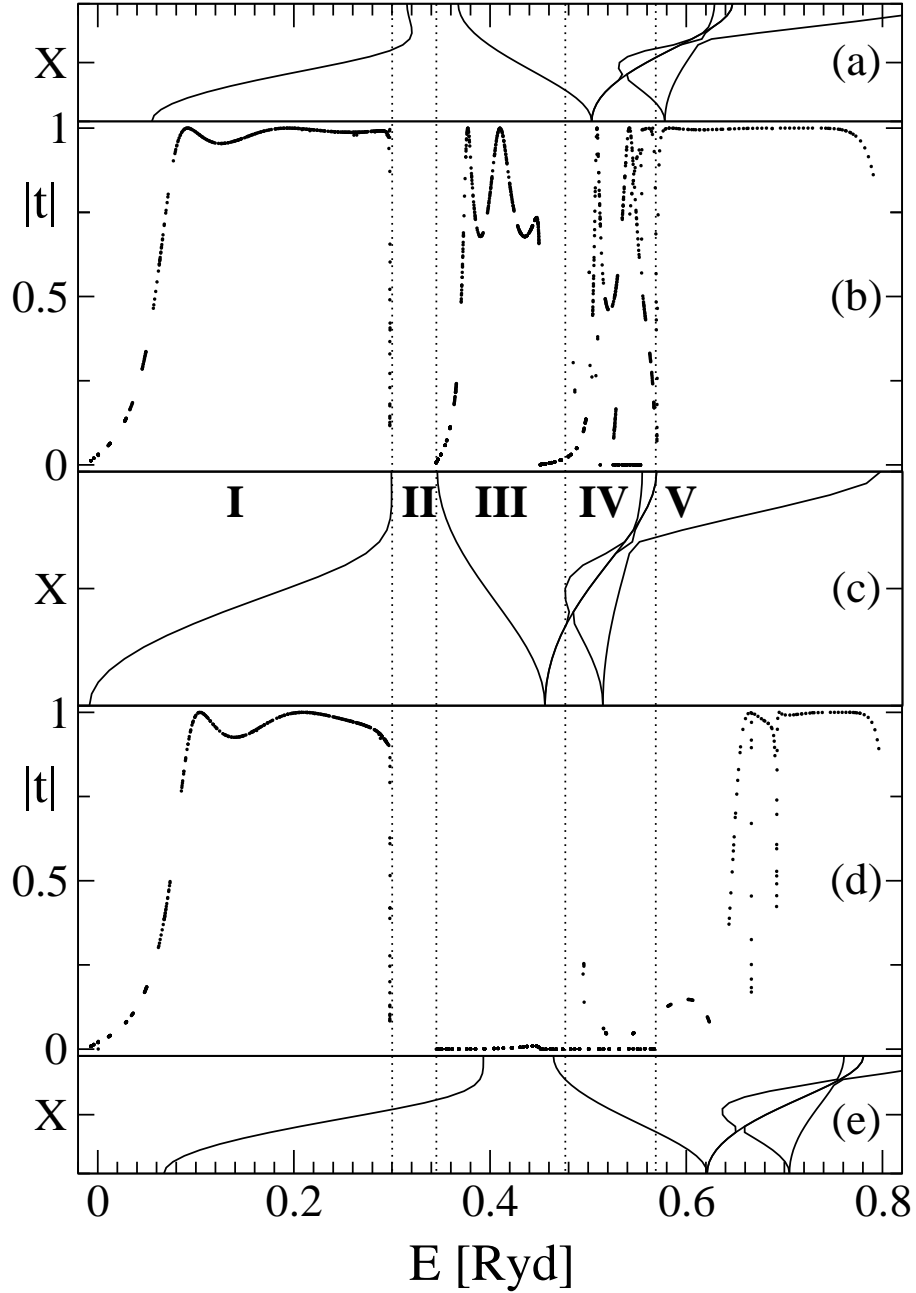}%
\end{center}
\vspace*{1cm}
  \caption{
Band structure of the majority (a) and minority (e) electrons of Co,
band structure of Cu (c), furthermore the $\left| t\right| (E)$-dependence
for Cu electrons through the majority (b) and minority (d) potential barrier
of 5 ML\ Co; all along the line $\Gamma-X$, that is ${\bf k}_{\parallel }=0$.
  }
\end{figure}
\newpage
%
\begin{figure}
\begin{center}
\parbox{1.1\Mlength}{
{\bf a)}  \includegraphics[width=\Mlength,clip=true
]{Fig6a.eps}\\
{\bf b)}  \includegraphics[width=\Mlength,clip=true
]{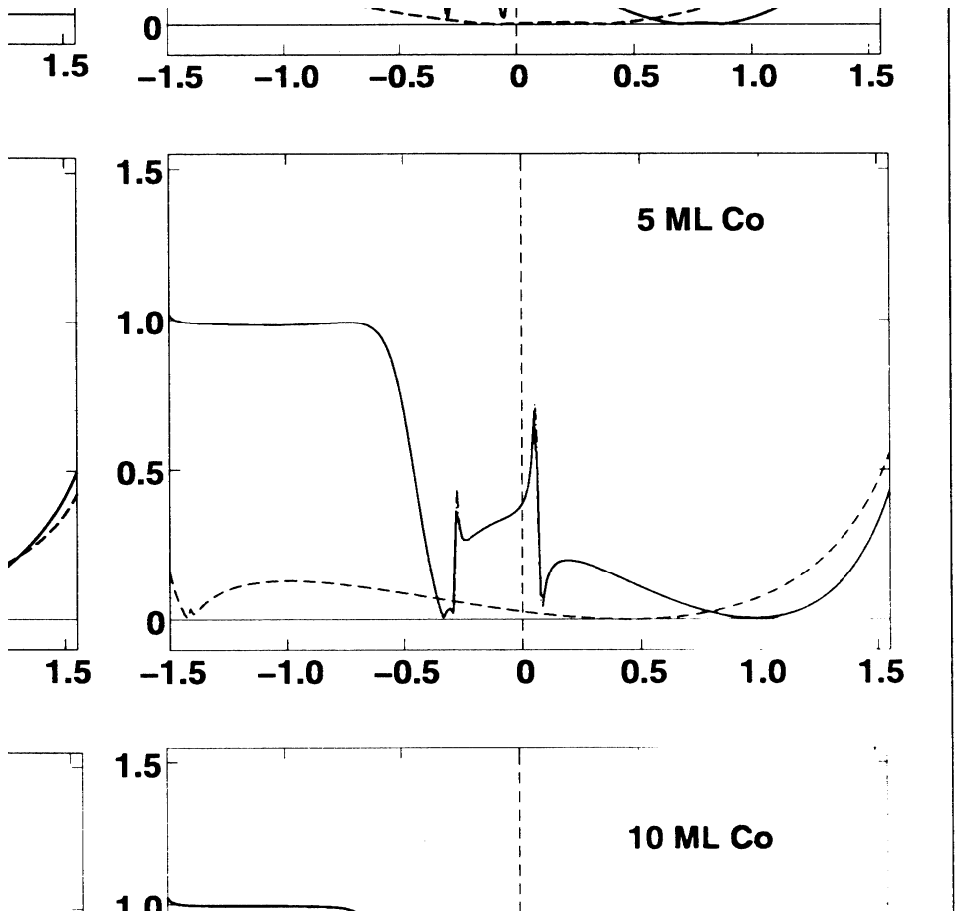}
}
\end{center}
\vspace*{1cm}
  \caption{
The dependence of $\left| r\right| $ on $(E-E_{F})$ with our
new method (a) for the majority (dashed line) and minority (solid line)
Cu electrons through 5 ML of Co(100) at ${\bf k}_{\parallel }=0$
in comparisson with Wildberger (b) (see Fig. 8.5, p.152 in
\protect\cite{c:Wildberger97}); the axes are the same for both figures.
  }
\end{figure}
\newpage
%
\begin{figure}
\begin{center}
  \includegraphics[width=\Mlength,clip=true]{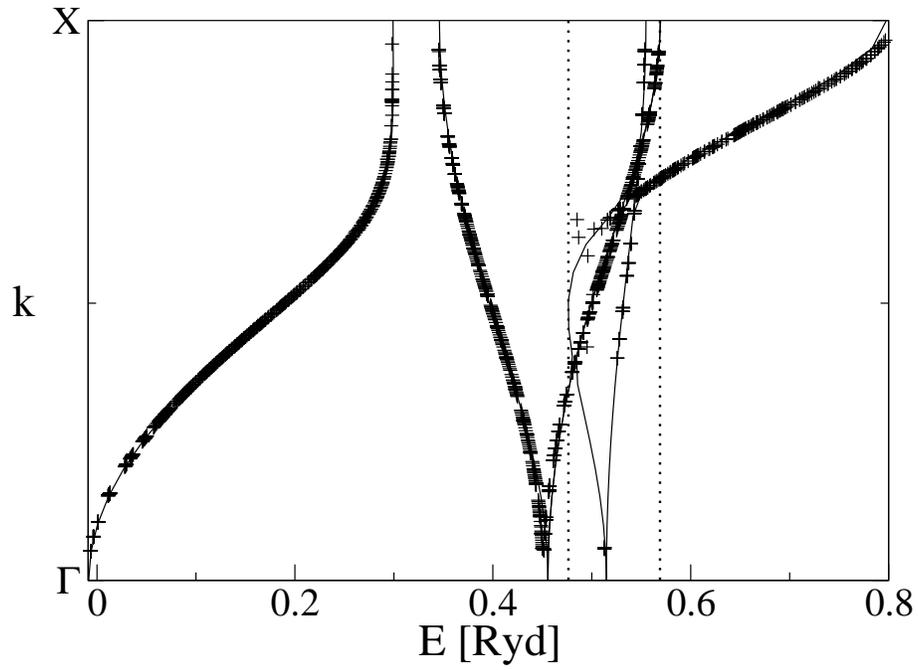}%
\end{center}
\vspace*{1cm}
  \caption{
Band structure of fcc Cu along $\Gamma X$ from a Cu bulk calculation
(solid line) and indirectly from the superlattice Cu/Co (001) calculation
(cross).
  }
\end{figure}
\end{document}